# A Journey in Shared Memory Land

**Ran Ginosar**

*The Andrew and Erna Viterbi Faculty of Electrical and Computer Engineering,*
*Technion—Israel Institute of Technology, Haifa 32000, Israel*

## 1. Summary

I have greatly enjoyed spending many years in studying parallel computing. My journey, detailed in the following pages, is summarized in Figure 1. As a PhD student at Princeton (1978—1982) I studied a combination of shared memory and message passing, motivated by algorithms and the ease of programming. As a young faculty member at the Technion (1980's), a brilliant MSc student figured a better way for shared memory machines that further simplifies parallel programming [1]. He later founded Plurality (2000's) and we developed the HAL machine [2]. At Ramon Chips (2010's), we reinvented RC64 [3] and managed to make programming much easier [4].

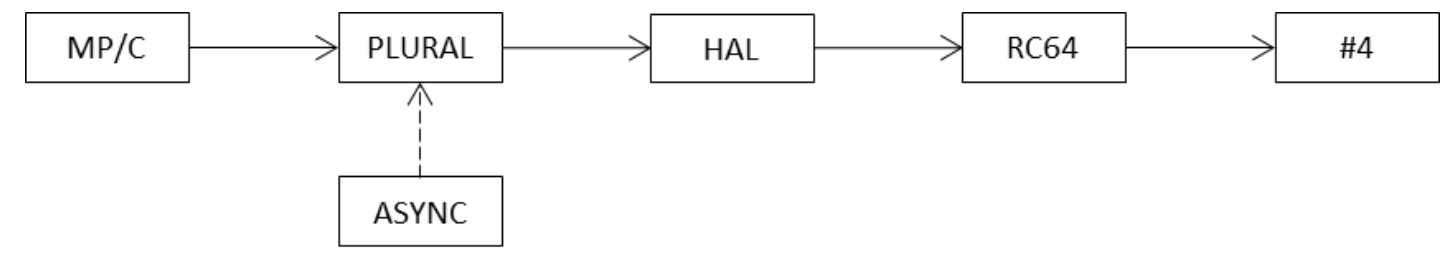

***Figure 1. My Parallel Journey***

Theoretical computing models, especially PRAM [5], have constituted the foundation for my work. Fortunately, I spent many years learning the theory and practice of asynchronous logic [6], and applied some of these concepts towards improving (namely, simplifying) parallel machines. Over the years, I became disillusioned of inventive hardware designers and their innovative machines. That perspective helped crystalize this process. Still, much more remains to be done!

## 2. MP/C

Parallel processing started to intrigue me as early as my sophomore year of computer engineering at the Technion (1975), once I began to realize what computing was all about. When I came to Princeton for my PhD in EECS (1978), I met a highly talented group of CS theoreticians who taught me to never blindly trust anything. I studied relational algebra and relational calculus in Jeff Ullman's database class [7], leading me to think about parallel processing in database machines. But Jeff moved to Stanford and, luckily, I got pulled by Bruce Arden to the mathematical foundations and performance evaluation of operating systems, computer systems, and networks. We studied Kleinrock's textbooks on queueing networks [8] and I started grasping the grand scale that opened the door to better understanding parallel and distributed computing. Seymour Cray came to town showing off his Cray-1 [9], and I realized it was overly complex, hard to program, and it was hard to envision what was going on within that machine. In some ways, he (and his colleagues at CDC [10]) anticipated modern GPUs, a successful breed that is still a nightmare to program [11]. With time, it has become evident that vector machines, including SIMD units and the (too many) variants of RISC-V vectors are hard to use, and vectorizing compilers are disappointing [12]. And chaining [13], another Cray novelty, has largely been forgotten.

*Author for correspondence (ran@ee.technion.ac.il).

While Manolis Katevenis carried the flag of simplicity in the CPU front at Berkeley (his work on RISC [14], inspired by John Cocke of IBM and David Patterson of Berkeley), I preferred investigating the parallel world. I checked Alessandro Zorat's work at USC on Divide-and-Conquer architectures [15], and thought that a better method might insert switches within the multi-master Multibus [16]. By 1979, we referred to a (multi-bank) shared memory system as a Multi-Processor, and the term Multi-Computer indicated separate processor-memory pairs (computers) exchanging messages over some interconnect, as shown in Figure 2 [17].

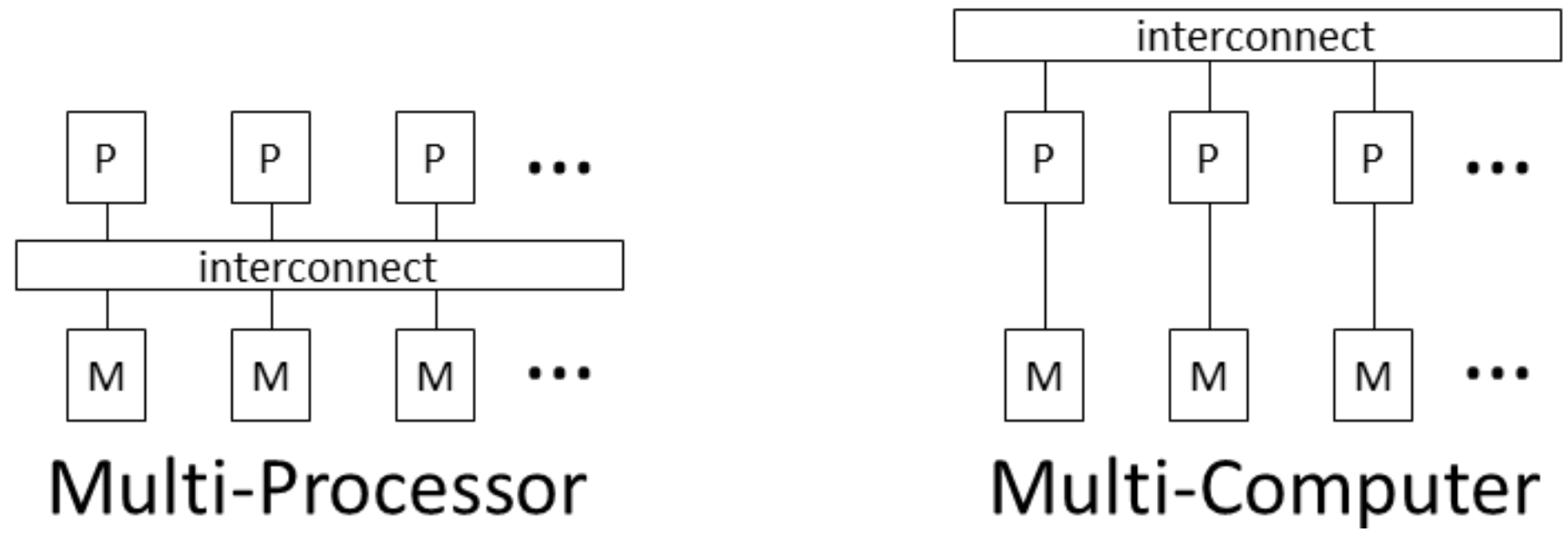

*Figure 2. MP and MC*

Inserting switches within the bus (see Figure 3) enabled divide-and-conquer algorithms in a seemingly simple manner. Consider a serial-parallel-serial algorithm. The serial code is executed on one processor that can access the entire memory (left side of Figure 4). Parallel code is enabled on the MC configuration (right side of Figure 4). The serial code is tasked with scattering data to ensure that each processor finds its inputs in its respective memory bank, gathering the results once the parallel code is complete.

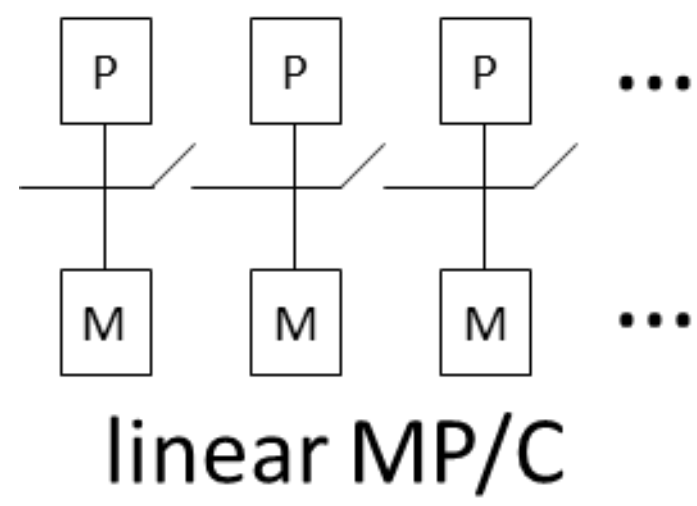

*Figure 3. MP/C can operate as MP when switches are closed and as MC when open*

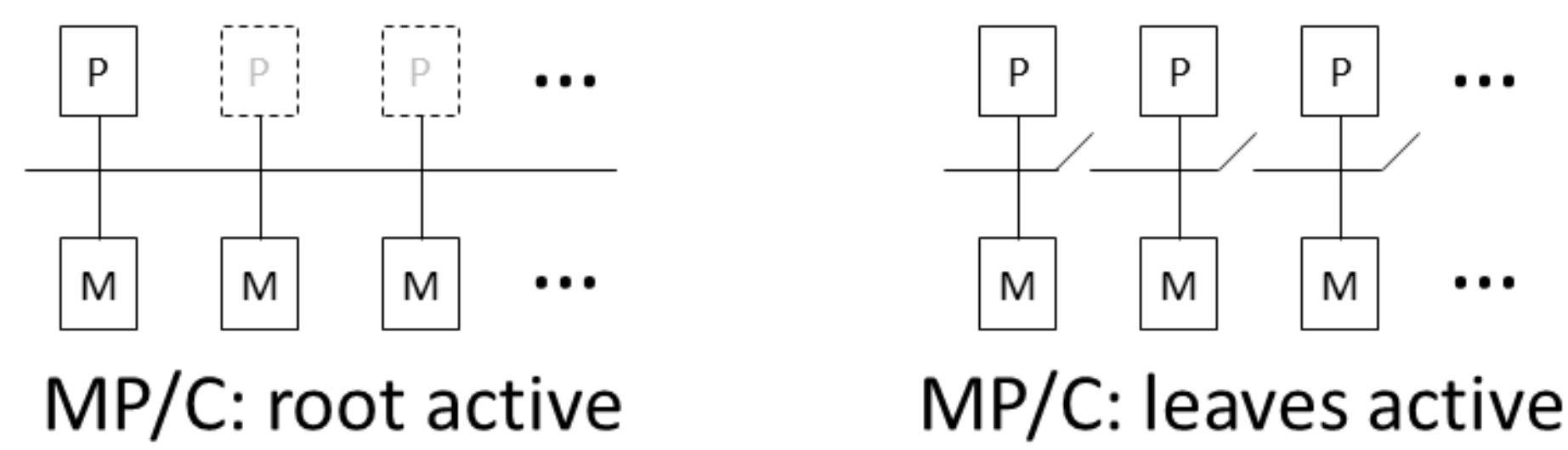

*Figure 4. Tree algorithms execute sibling nodes in parallel without interference*

MP/C faced several challenges. First, memory banks are of predetermined size, not great for dynamic allocation. Second, the algorithm had to manage *scatter* and *gather* activities [18]—making the code architecture-dependent, counter to seeking simplicity. I realized that, for elegant and efficient parallel processing, we had to separate the algorithm from the data movements. Third, the code had to manage switches, either as pragmas or as function calls.

Upon graduation, I turned down the idea to work with Jack Dennis at MIT on dataflow, an appealing architecture that seemed to produce unpredictable and data-dependent data movements [19]. I also stayed away from functional programming, a descendent of dataflow that employed inefficient control to manage run-time parallelism. Systolic arrays [20] didn't attract me either; while data movements were more predictable than in dataflow machines, the idea of continuously moving all data appeared strange for two reasons. First, systolic flow of blood feeds hungry body cells; should a systolic flow of data feed hungry processors? I preferred processors serving data rather than data serving processors. Second, systolic implied strict synchrony—no flexibility in latency nor in flow topology, severely limiting applicability. Indeed, these two reasons continue to haunt all systolic array based machine learning accelerators of today. Gene Amdahl asked me to join his IBM-clone company, but I realized I would end up in engineering and microarchitecture rather than be able to play architect. I decided to go to Bell Labs, which granted me full freedom to conduct my own research. I joined the glorious department where C, Unix and Shell were being developed.

At Bell Labs, in 1982—1983, I designed and fabricated bus switching chips (at their Allentown fab—BTL were among the first to offer VLSI shuttle to its own scientists), and ported Tom London's BTL Vax Unix (the basis of Berkeley Unix) to the SUN multi-processor system [21] that combined multiple Motorola 68000 CISC processors, multiple memory banks and a multi-master bus. The SUN system was originally made by Andy Bechtolsheim as a node on the Stanford University Network (SUN), which later formed the seed of Sun Microsystems. My Unix port to SUN system supported bus switching, and emulated MP/C. The experience helped me teach parallel computing when I moved back to the Technion by the end of 1983. As a farewell gift from BTL, the VP and Nobel laureate Arno Penzias gave me his Big Bang T-shirt! [22]

# 3. PRAM

While still at Princeton, I became acquainted with and fascinated by the elegant theoretic model of parallel computing. PRAM [5] is the parallel descendant of RAM, Random Access Machine [23], the all-time highly successful model of computing that forms the base of computer science. While RAM assumes that at each time unit a processor either performs one computing step or accesses one memory location, PRAM (Figure 5) states that at each time unit multiple processors can each either execute a single computing step or access a memory location. The beauty lies in parallel memory access—many processors can all access memory at the same time in one single step. CS theoreticians including Uzi Vishkin at the Technion [24] figured out how to program such a theoretical model, in manners that were not too different than the code I made for MP/C. Still, I found the vanilla PRAM model overly synchronized, in a SIMD way, because PRAM code appeared to require that all computing cores at any one step execute the same code statement.

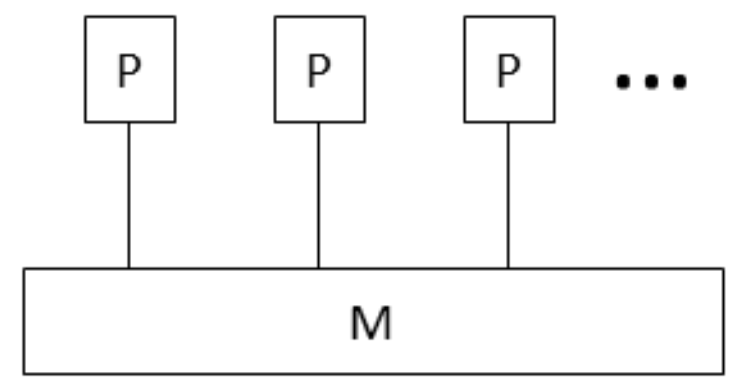

*Figure 5. PRAM, Parallel Random Access Machine model*

In 1980, Jack Schwartz at NYU inserted a logarithmic multistage interconnection network at the heart of the PRAM model (Ultracomputer [25]in Figure 6). Unlike the perceived SIMD character of PRAM, Ultracomputer adopted a MIMD view in which different processors can execute different programs, and only shared memory access was controlled. His colleagues at NYU broke down the shared memory into many memory banks and introduced the memory access conflict resolving Fetch & Add operation [26] that executed in a distributed fashion on the log-net. They constructed the NYU Ultracomputer (Figure 6, right), followed by the IBM Yorktown RP3 experimental machine [27]. The Fetch & Add (later, Fetch & Op [28]) trick was intriguing, but seemed overly complex and the machine required its own language and compiler. I used to teach it in my parallel computing classes but was not tempted to follow in its tracks.

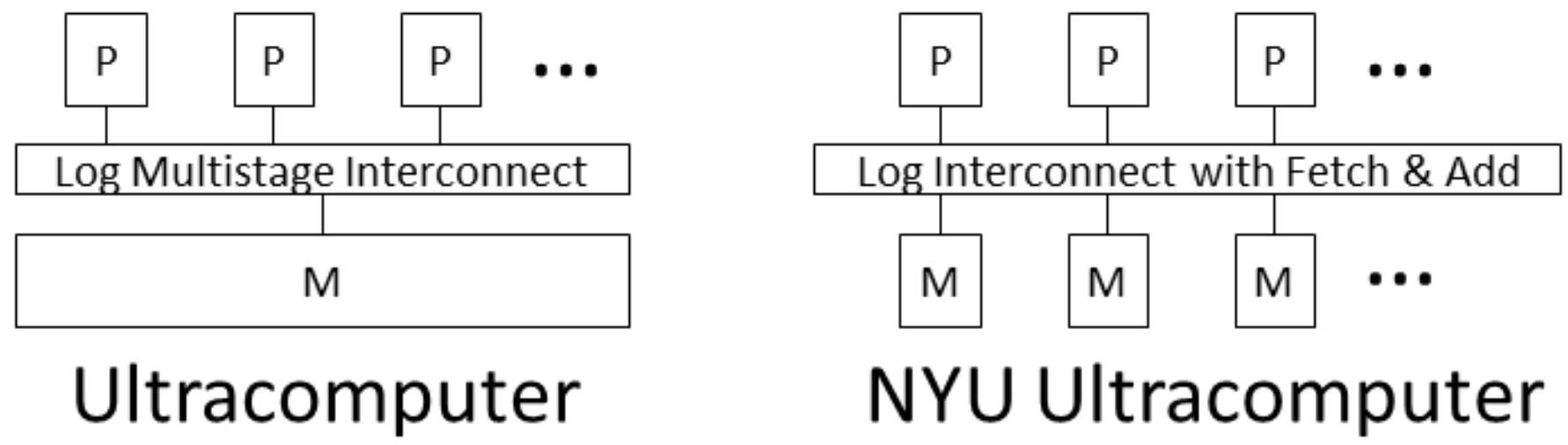

***Figure 6. Theoretical and "practical" ultracomputer models***

Marc Snir, one of the contributors at NYU and the external reader of my dissertation, ended up developing MPI [29], the message passing interface, yet another direction that appeared to me to be overly complex. Was Marc disillusioned of shared memory by his experience with NYU ultra? A related research camp developed message passing multi-computers using different types of interconnection networks, from hypercubes to 2D meshes [30]. Over the years, this architecture has proven quite inefficient, as demonstrated by the steep decline of the Tilera mesh-based manycore from presumptuous general purpose machine down to the packet-processing Bluefield (where the interconnect doesn't really matter) [31], and recognizing the inefficiency of the ring-based Xeon-Phy [32].

In sharp contrast with elegant PRAM derivatives, multicomputers at the time (and multicores of later eras) were modeled as alternating series-parallel phases, inspired by a misguided application of Amdahl's law [33][34]. That approach led to programming languages using explicit parallelism (e.g., parallel-do blocks), eventually leading to obscure languages such as CUDA [35]. Other languages, such as Cilk [36], enabled run-time data-dependent spawning of nested parallelism, creating complex structures that were difficult to envision, control, verify, and efficiently apply. I realized that I should stay close to the more foundational PRAM model.

## 4. PLURAL

Two big leaps forward occured in the late 1980's, hoping to address the complexities of the aforementioned PRAM approximations. Uzi Vishkin converted from pure math to computer architecture and introduced XMT (Figure 7, left) [37]. Rather than F&A, prefix sum was employed for more general management of simultaneous access to memory by many processors. Rather than scheduling code segments to processors by software, a hardware scheduler was introduced. The schedular also used prefix sum, to enable simultaneous issue of code segments to the many processors. XMT expanded and elaborated on the PRAM model. It was highly efficient in programming and executing challenging parallel structures such as graph algorithms, which invoked code segments in data-dependent and unpredictable manners.

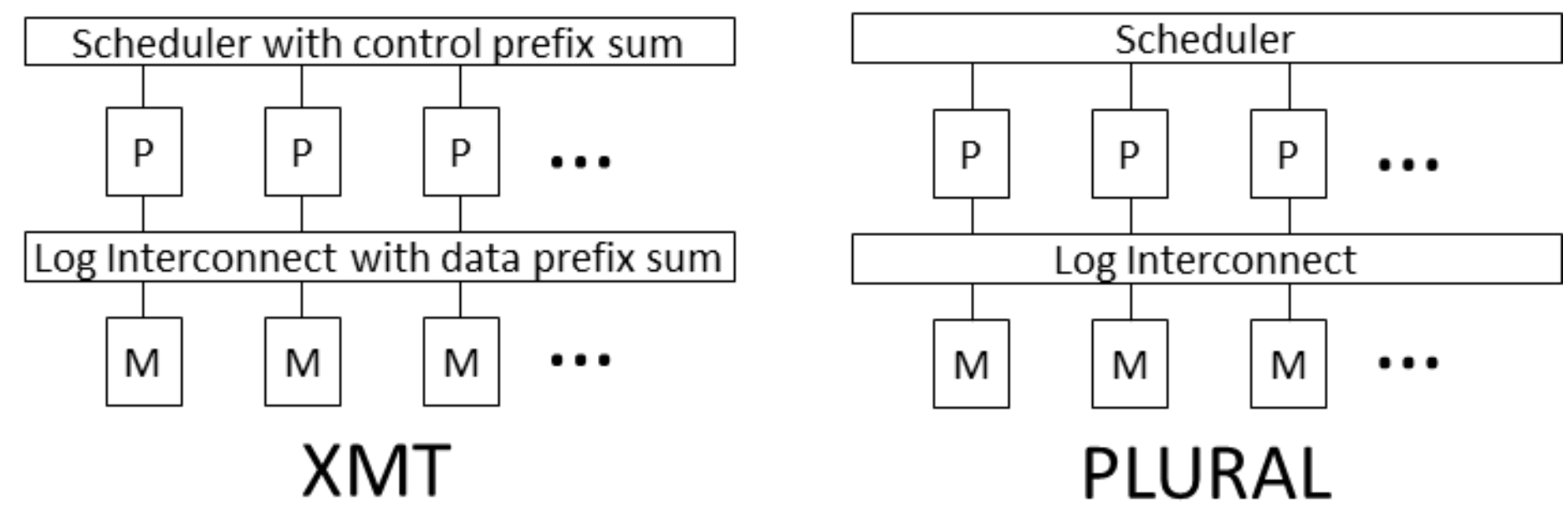

***Figure 7. Scheduled shared memory manycores***

Simultaneously, and totally independent of Uzi's work, my MSc student Nimrod Bayer introduced PLURAL, a shared memory manycore architecture (Figure 7, right) [2]. Similar to XMT (and NYU ultracomputer), PLURAL used a log-net as a *bipartite* connection of multiple processors to multiple memory banks. Similar to XMT, PLURAL included a hardware scheduler. But everything else in PLURAL was quite different from XMT. The line of argument that led to PLURAL was as follows:

- A PRAM-inspired MIMD model was envisioned.
- The Concurrent-Read-Exclusive-Write (CREW) variant [38] of the PRAM model was adopted.
- The wisdom of compiler theory that used graphs to represent program flow, precedence relations and dependencies of code blocks was applied.
- Data dependencies (which burdened dataflow architectures and functional programming) were translated into equivalent dependencies among code blocks. We noted that such clean translations required shared memory, and were much harder to manage in message passing "MC" architectures.
- The set of all translated dependencies among code blocks made up a directed dependency graph.
- Consequently, programs were divided into two parts: a set of serial code blocks, renamed *tasks*, and a graph identifying task inter-dependencies. Thus, the *Task Oriented Programming Model* was born[1]. We realized that the model was useful for simple data-parallel algorithms (such as linear algebra and signal processing) but (unlike XMT) difficult to apply to complex data dependent, graph, and recursive algorithms.
- Neither hacking, nor assembly level programming, nor any special cases were allowed. All programming was forced to strictly abide by the rules of the task oriented programming model. The model inspired programmers to generate code that was 'correct by construction'. Years later, this model paved the road to formal definitions and formal verification of every PLURAL program.
- A hardware synchronizer/scheduler was devised to 'execute' the task dependency graph, manage the state of dependencies, schedule tasks for execution, and collect completion notices.
- Tasks were compiled using standard compilers, and the task dependency graph was converted for execution by the scheduler.
- Some tasks were defined 'regular' and could generate completion codes that would affect branching decision making within the graph. Other tasks were 'duplicable' and would be duplicated into a pre-determined number of concurrent (hence mutually independent) instances.
- When all processors execute instances of a single duplicable tasks, the manycore follows the SPMD discipline. Note that multiple independent (hence concurrent) duplicable tasks could be simultaneously executed, enabling highly complex multiple-SPMD scenarios.
- No data movements were needed between task executions. All data (variables) were stored in shared memory and made available to all tasks according to the PRAM model.

The PLURAL architecture constituted no less than a breakthrough. It helped us get closer to the ideal world of easy programming that should result in efficient parallel execution. But PLURAL would take many more years to be demonstrated in practice.

# 5. Plural Programming Model

The Plural PRAM-like programming model [3] is based on non-preemptive execution of multiple sequential tasks. The programmer defines the tasks, as well as their dependencies and priorities which are specified by a (directed) *task graph*. The tasks are executed by cores and the task graph is 'executed' by the scheduler.

In the Plural shared-memory programming model, concurrent tasks cannot inter-communicate. A group of tasks that are allowed to execute in parallel may share read-only data, but cannot share data written by any one of them. If one (*producer*) task must write into a shared data variable and another (*consumer*) task must read that data, they are *dependent*—the writing producer task must complete before the reading consumer task may commence. This dependency is specified as a directed edge in the task graph, and is enforced by the hardware scheduler. Tasks that do not write-share data are defined as *independent*, and may execute concurrently. Concurrent execution does not necessarily happen at the same time—concurrent tasks may execute together or at any order, as determined by the scheduler.

[1] In other literary circles this is named *Task-Based Programming* [39]

Some tasks, typically amenable to independent data parallelism, may be *duplicable*, accompanied by a *quota* that determines the number of instances that should be executed (declared parallelism). All instances of the same duplicable task are mutually independent (do not write-share any data) and concurrent, and hence may be executed in parallel or in any arbitrary order. These instances are distinguishable from each other merely by their *instance number*. Ideally, their execution time is short (fine granularity). Concurrent instances can be scheduled for execution in any (arbitrary) order, and no priority is associated with instances.

Each task progresses through at most four states (Figure 8). Tasks without predecessors (enabled at the beginning of program execution) start in the *ready* state. Tasks that depend on predecessor tasks start in the *pending* state. Once all predecessors to a task have completed, the task becomes *ready,* and the scheduler may schedule its instances for execution and allocate (dispatch) the instances to cores. Once all instances of a task have been allocated, the task is *All allocated.* And once all its instances have terminated, the task moves into the *terminated* state (possibly enabling its successor tasks to become *ready*).

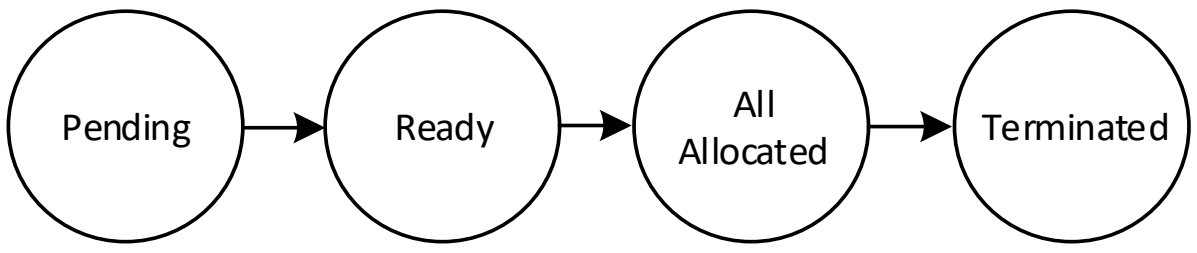

***Figure 8. Task State Graph***

Data dependencies are expressed (by the programmer) as task dependencies. For instance, if a variable is written by task $t_w$ and is later read, reading must occur in a group of tasks $\{t_r\}$ and $t_w \rightarrow \{t_r\}$. The synchronization action of completion of $t_w$ prior to any execution of tasks $\{t_r\}$ provides the needed barrier.

The task graph, as well as the code, is fixed at compile time. No run time code or graph changes are permitted. The only parameter that can be passed from the code to the task graph is the quota of a duplicable task, and even that parameter cannot depend on data values to avoid recursion and run-away parallelism.

# 6. Async

I have fortunately been able to share some insights from the art and science of asynchronous logic design to my work on parallel processing.

Chuck Seitz' chapter 7 of Mead & Conway VLSI book [40] got me fascinated by metastability, synchronizers, and the hope of removing clocks. Michael Yoeli introduced me to temporal logic and mathematics based design [41][42], and I greatly enjoyed taking it to higher levels by learning from Alain Martin [43] and Jo Ebergen [44]. However, the purist, and even quasi-purist, "delay-insensitive" (DI) approaches turned out to be too costly. These ideas did lead to high performance circuits by Fulcrum Technology and Peter Beerel [45] and I have designed several DI processors as academic exercises. In the end, DI methods did not appear to be attractive for parallel processing.

The other camp adopted relative timing and applied a different type of mathematics to the design of async circuits. I followed studies by the Victor Varshavsky school, Luciano Lavagno and Jordi Cortadella [46], who adopted mathematical management tools based on graph theory. I was impressed by the use of Petri nets and signal transition graphs, and by the ability to define properties that guide synthesis algorithms and methods for formal verification. At Intel, with Ken Stevens and others, we applied the approach to the RAPPID design [46][47]. I taught that art at the Technion for several years, and soon realized that the theory and tools would be useful in general for distributed and parallel processing, and would be particularly applicable to task graphs. That realization helped me stick with task graphs as key to efficient parallelism, and avoid all sorts of parallel programming languages. This realization also forms the basis for Section 9 below on formal verification of RC64 programs.

Steve Furber led efforts on applying self-timed methods to processor design [49]. Rakefet Kol followed with a different version [50]. Eventually, Steve and the Amulet adventure helped me realize that, except for extreme cases of very high performance or very low power and energy, asynchronous design introduces benefits in education, mathematics, and formal and theoretical treatments, but is less attractive as a practical engineering method.

While asynchronous processors did not revolutionize the field, the lessons of the asynchronous design theory do apply to -chip interconnects of many cores, memories and other entities. Reuven Dobkin developed an asynchronous NoC router [51].and together with Christos Sotiriou we have studied GALS architectures [52], further discussed in Section 12 below.

# 7. Plurality HAL

The PLURAL architecture was developed with mainframes and supercomputers in mind. In the early 1990's, no large mainframe manufacturer was interested, the architecture therefore got shelved, and Nimrod turned to get his PhD in mathematics. But a decade later, silicon integration matured to where we could imagine a few processors and a few memory banks integrated onto a single chip. Nimrod and I turned to Intel in 2003 but they kicked us out, saying that parallel processing was not competitive, and that Intel could achieve better speedup by repeatedly porting their great Pentium family processors to the next silicon generation. A year later, Intel introduced the multicore. By the way, that seminal move by Intel motivated us and others to adopt a new term, *manycore*, for our architectures. Ever since, we have had to explain that "manycore is not a multicore." Eventually, Nimrod managed to raise funds and founded Plurality.

At Plurality we reduced PLURAL to silicon, named HAL (formally HyperCore Architecture Line, in loving memory of Clarke's Space Odyssey HAL, the Heuristically programmed ALgorithmic computer) [2]. A conceptual description of the architecture is shown in Figure 9.

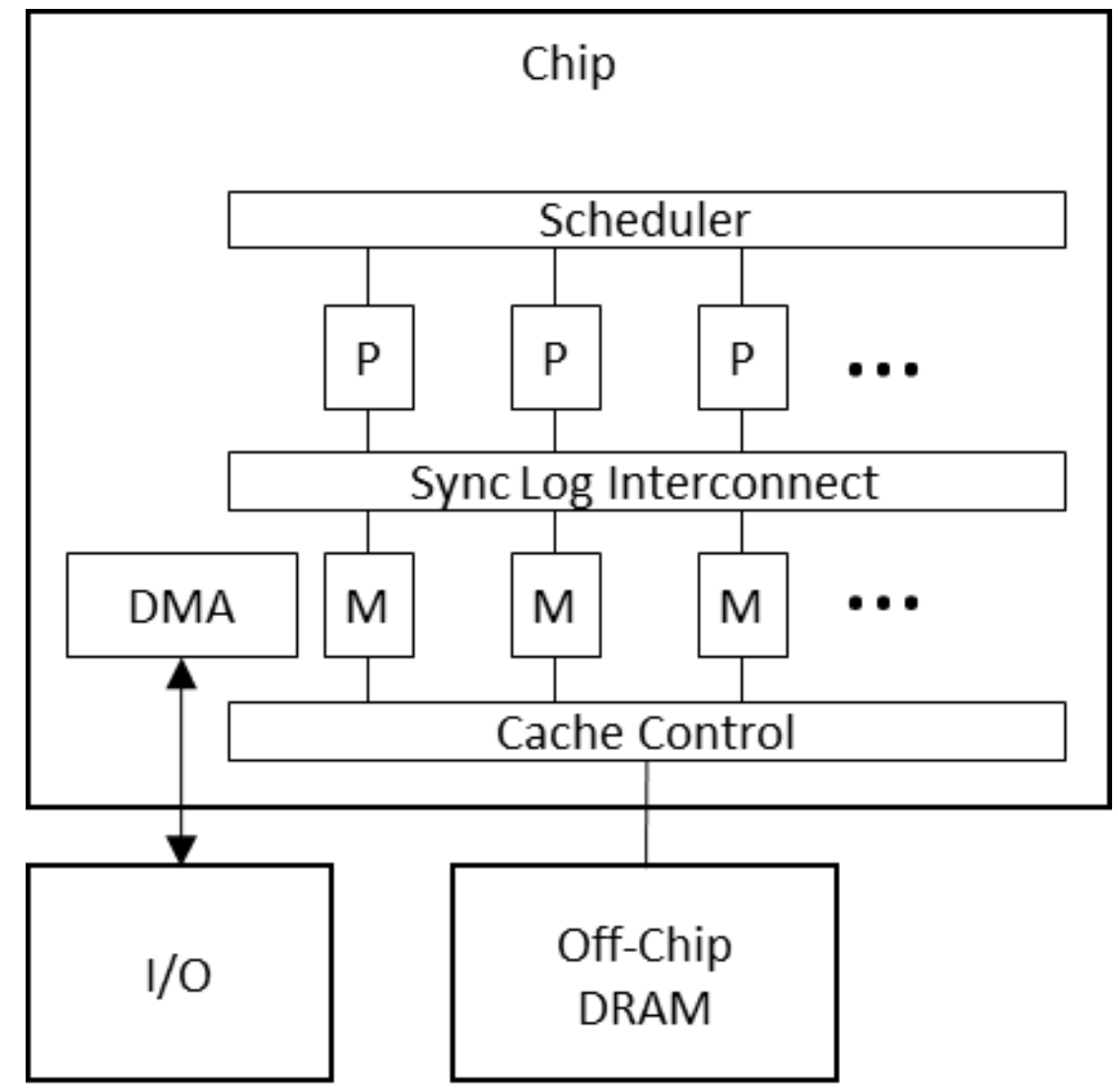

***Figure 9. HAL from Plurality was the first VLSI implementation of PLURAL***

Working with silicon motivated us to develop the following features, relative to PLURAL:

- Processors were designed as memory-less "cores," using commercially available IP cores. We considered several different cores, and designed the rest of the architecture in a core agnostic manner. All load and store instructions accessed the on-chip shared memory.
- The on-chip shared memory, perceived as too small, was managed as a shared cache, acting as a front-end to an external DRAM. This choice was similar to last level cache in modern multi-cores. In retrospect, it was difficult to manage cache-aware programming in parallel computing, and HAL experienced thrashing at times.
- The logarithmic multistage network connecting cores and memory (cache) banks was designed as a synchronous circuit in an effort to ensure fixed latency through the net. The network included input and output registers, and the entire path through the network was designed as a combinational circuit. This design efficiently avoided a large number of pipeline registers in between the stages within the network.

- In case of conflicts in memory access operations, whether they happen in the lognet switches or at bank ports, one of the conflicting access operations was aborted. The aborted core received a NACK and had to re-issue the access operation.
- Once a core completes executing a task, it is available for executing any other *ready* task, since the cores do not hold state.
- All I/O was managed by DMA access directly to and from the memory banks.

Plurality HAL was initially built on a FPGA, followed by an eASIC 'structured silicon' version. A complete software stack was created, and several applications were demonstrated. While PLURAL was perceived as a linear algebra accelerator, we managed to show efficient lossless and lossy compression, exploiting the ability to program and execute multiple levels of parallelism.

Consider JPEG image compression shown in Figure 10. Each step can be parallelized, but the steps must be executed in order, limiting the utilization of the 64-core HAL machine. To overcome this limitation, a 'soft' pipeline is created by means of the task graph shown in Figure 11. Each algorithm step is applied to a different set of image blocks. The many steps shown in the graph are concurrent, namely are independent of each other and may all run simultaneously, or at any order. Once certain cores have completed executing a step, these core are available to execute other steps that may be waiting, as determined solely by the scheduler. We name such task graphs *manyflow*. Unlike hardware pipeline, the pipe stages may execute at different times rather than simultaneously, and there is no need to balance the pipe.

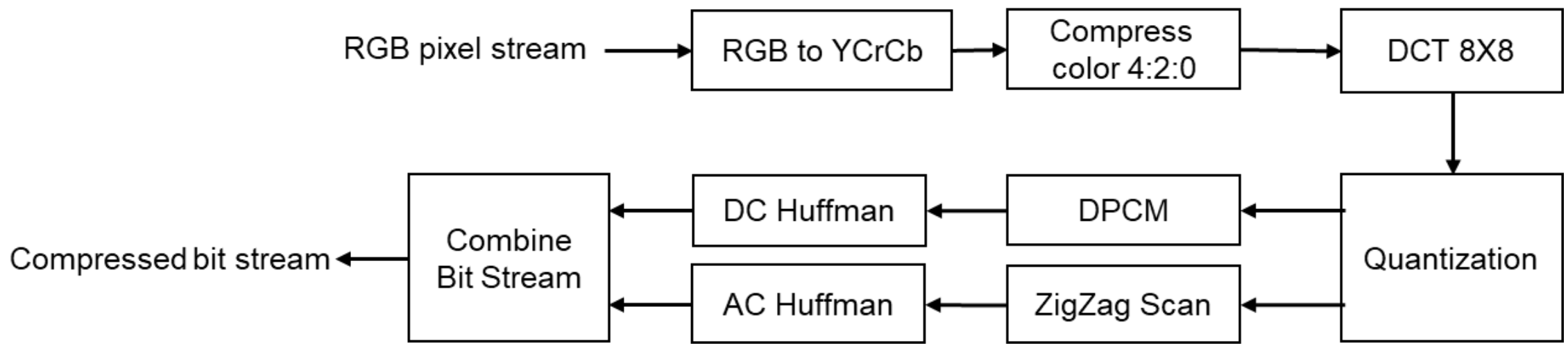

*Figure 10. JPEG image compression algorithm*

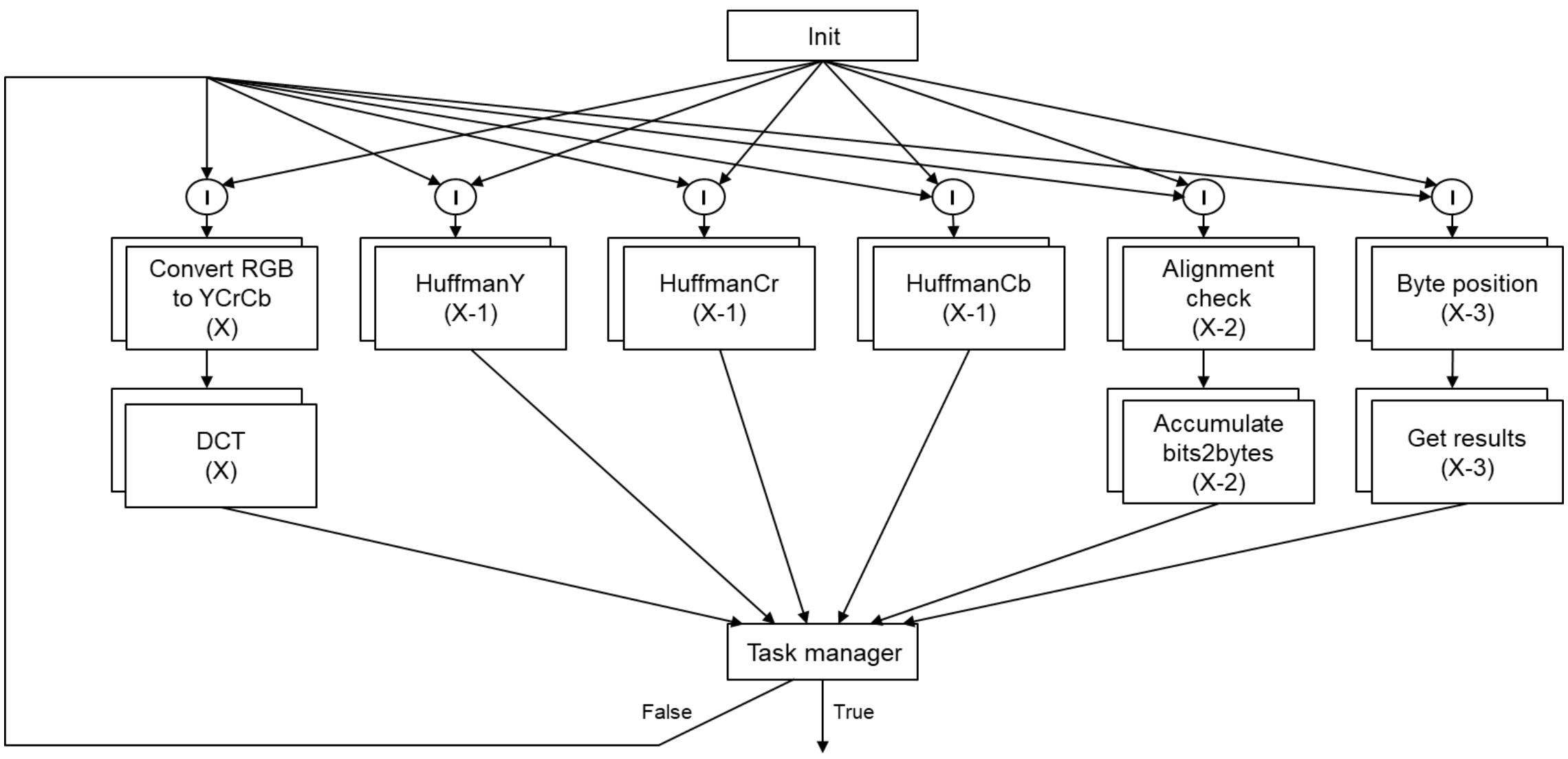

***Figure 11. Manyflow task graph for pipelined execution of JPEG. Duplicable tasks are shown as shaded blocks. The Task manager block is not a computational task and is not allocated to any processor—it is merely a loop control within the graph, executed solely by the scheduler. "X" indicates input generation.***

The best way we found for demonstrating performance, assessing power, and facilitating code development and debugging was based on charts that illustrated how many cores were busy at any time, down to the cycle-by-cycle level. Utilization by task is shown in Figure 12, and overall performance is shown in Figure 13. A similar chart (not shown) reports the instantaneous level of power consumed at each cycle. It is evident that manyflow programming can achieve high performance on a complex algorithm that is extremely difficult to parallelize using conventional approaches. The same lesson served us well when addressing other challenges, including telecom modems and SSD controllers.

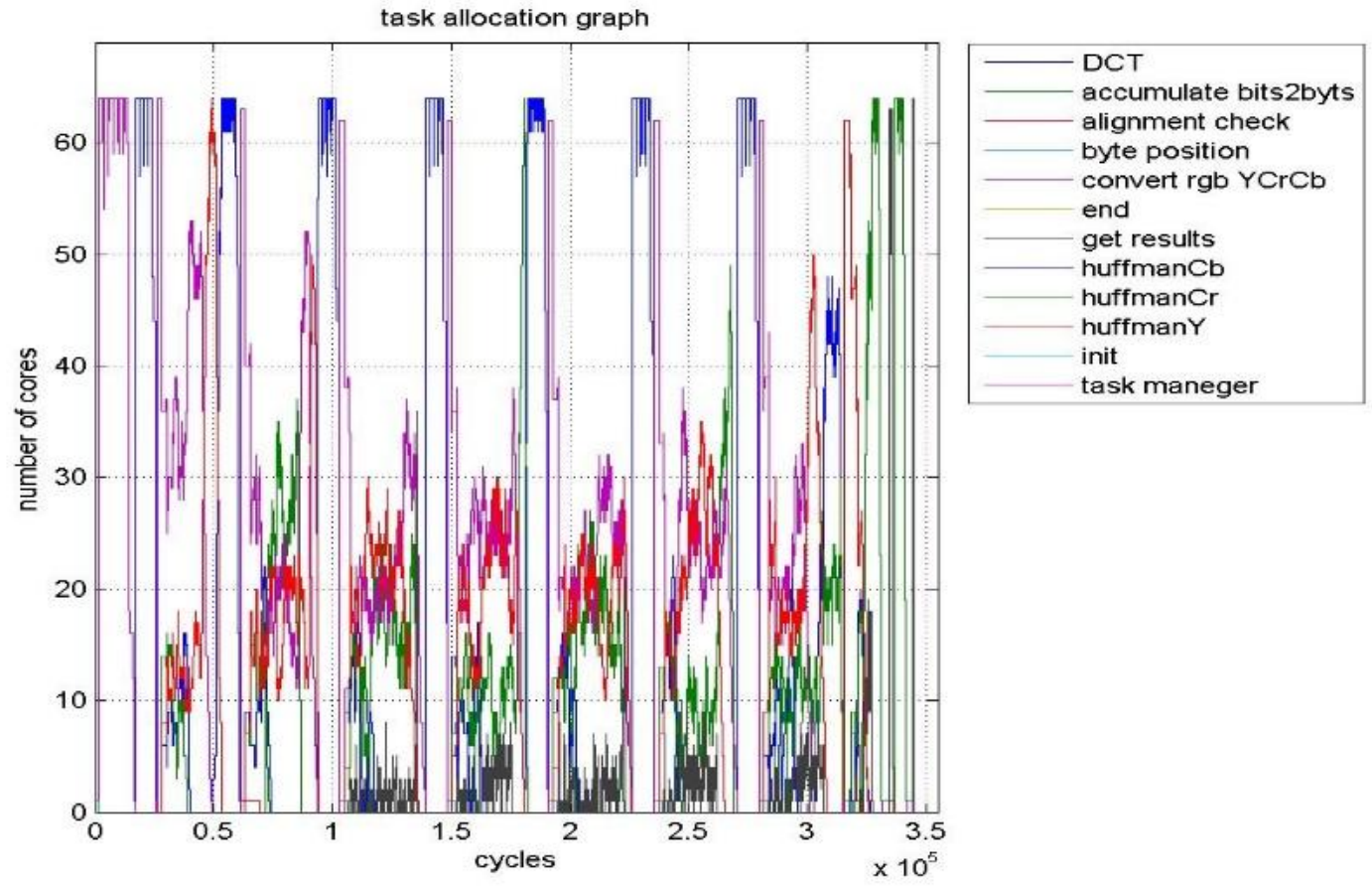

*Figure 12. HAL performance expressed as core utilization. The horizontal axis shows time (clock cycles), and the vertical axis indicates number of busy cores, up to 64. Each task is represented by a chart. While useful to programmers during development, it is difficult to envision overall utilization.*

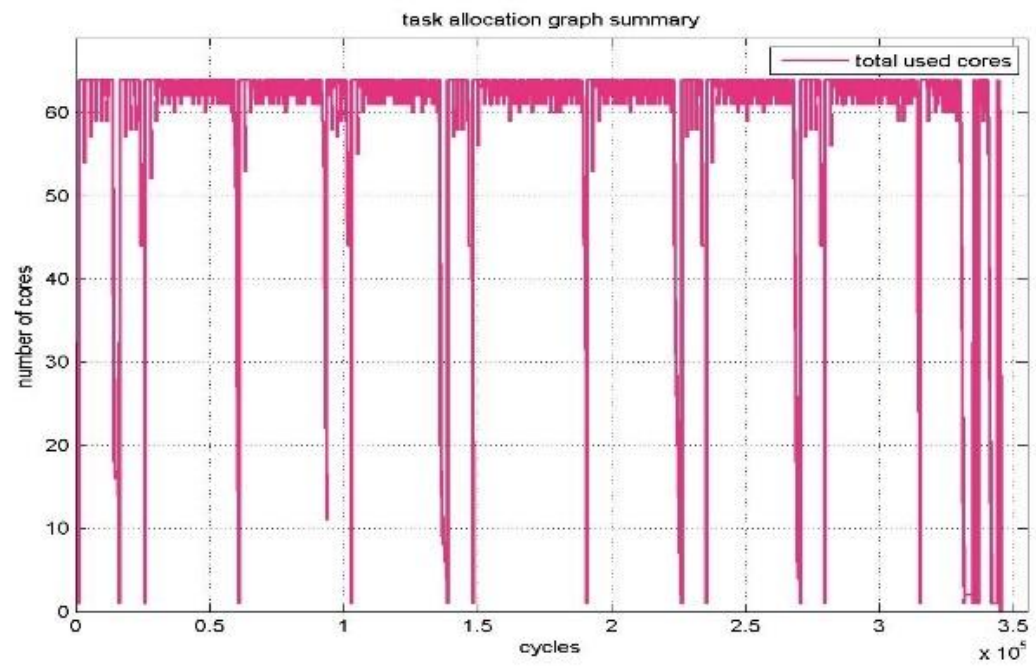

*Figure 13. Overall JPEG execution achieves higher than 95% core utilization. The red vertical lines indicate synchronization points, when all concurrent tasks have completed, and the scheduler is initiating the next phase of the pipeline.*

## 8. Ramon Chips RC64

The Plurality startup performed well on the technical side and poorly on the business side. Plurality closed its doors in 2011, but the architectural concepts did not disappear. I co-founded Ramon Chips (now Ramon.Space) back in 2004 to make radiation hardened processors for use in Space missions, and that company was ready by 2013 to engage in high performance manycores for Space applications. The Space community happily adopted my proposal to base the next generation of Space DSP on PLURAL and HAL architectures, and we received generous funding from the Israel Space Agency and the European Commission FP7 / Horizon programs. The design was named RC64 (the initials of the company name and 64 cores) [3]. RC64 (Figure 14), fabricated in 2017 in the TSMC 65nm low power process, incorporated significant improvements over HAL, as follows.

*Figure 14. RC64*

- The memory-less cores of HAL were replaced by DSP cores (from CEVA) featuring VLIW and SIMD. The cores can be programmed by plain C code and function as RISC cores, dissipating significantly less power than fully utilizing all SIMD MACs.
- The cores are power efficient, using a low voltage and operating at a low clock frequency. Overall performance is achieved by high levels of parallelism rather than by sheer speed.
- A floating point FMA vector unit is appended to each core
- An encoder/decoder (for LDPC code), marked as an accelerator in Figure 14, is added at the SoC level, connected to shared memory through the log net, and placed on the same side of the log net as the cores.
- Each core is equipped with three private memories: an instruction cache, a write-through data cache, and a private memory.
- The data caches are invalidated at the end of each task execution, to ensure coherency and avoid incidental cache reuse across task boundaries.
- The private memories house the C-language stack. This feature enables elegant parallel processing, as explained below.
- The shared memory was increased to 4MB (6MB including ECC), and was detached from external memories. The use of shared memory as a cache was eliminated.
- Shared memory is organized in 256 banks, four times the number of cores, to enable simultaneous access by the many cores. To reduce collisions, the addresses are interleaved over the banks.
- There is no cache coherency mechanism.
- The logarithmic network connecting cores to the memory banks is pipelined, upgraded from the combinational circuits within HAL.
- The I/O are managed by DMA to/from shared memory banks, through the log net. Thus, the DMA controllers (aka IO processors) are treated similarly to the processing cores.
- Task priority is determined by the order of rows in the text file describing the task graph. Priority ensures allocating cores to regular tasks prior to flooding the cores with instances of duplicable tasks, to prevent execution bubbles and to enhance utilization.
- Only one duplicable task may be scheduled at a time. Once all instances of a task are scheduled, instances of another ready task can be scheduled, even if some instances of the previous task are still executing. This approach simplifies the scheduler, while not compromising the efficiency of complex algorithms such as JPEG.

- The tasks are expressed and handled as C-language functions, enabling the use of standard compilers and avoiding unique languages.
- The task graph model is extended to include an interface to external (asynchronous) events such as interrupts and completion of DMA steps. DMA controllers issue pre-allocated *hardware event tokens* to the scheduler. A novel type of high priority task is defined, which is enabled by event tokens. Once enabled, the high priority task preempts execution of one core (without a complete context switch), performs short interrupt handling jobs, and releases the preempted task or instance. Since there are more cores than DMA controllers and interrupt sources, there is no need for handling nested interrupts and no need for masking interrupts. We note that while hardware event tokens and high priority tasks contribute to efficient handling of I/O, those mechanisms can be disruptive in the wrong hands. Hence, only system programmers are granted permission to use those mechanisms, and they are unavailable to algorithm programmers.

RC64 came out as an elegant and efficient manycore. It turned out highly effective for both simple (regular linear algebra) and complex algorithms [57][4]. RC64 has demonstrated high efficiency on deep learning, readily accommodating large convolution tensors as well as nonlinear pooling and other kernels [53]-[56]. It has excelled in high reliability SSD controllers, where many reactive tasks are tightly coupled with high rates of I/O operations, and where several unrelated flows of data take place concurrently. The aforementioned programming model and performance evaluation tools facilitate effective tuning during code development.

Programming frameworks and languages such as Matlab and OpenMP may be compiled into a RC64 C-based task oriented programming model. However, to fully exploit RC64 flexible parallelism, strict constraints need to be imposed on programs written in these frameworks and languages. These constraints and restrictions actually eliminated useless 'fluff' that had been introduced into such languages.

The RC64 architecture, like PLURAL and HAL, thrives on fine granularity. The shorter the tasks and instances, the easier to balance the load and approach 100% utilization. Several factors, however, incur overhead and pose an effective lower bound on grain size. Tasks go through prologue and epilogue sections of C functions. Each task starts with a cold cache and may be delayed by cache misses. The scheduler may also be delayed in responding to task completion notices. Consequently, part of code tuning requires searching for optimal granularity (typically, more than 1000 cycles long). In the next generation (see below), advanced scheduling and pre-fetching are considered in response to this issue.

# 9. Formal Verification

The programming model is designed to enable two types of formal verification. The task graph should be verified for properties that assure correct concurrency, such as safety (at most one token per edge, tokens don't overtake each other), liveness (no deadlocks), and persistence (ready tasks eventually execute). The combined code and graph should be verified that the CREW rules are not violated, to assure memory consistency and correct PRAM-like execution.

A clear distinction is made among the terms parallel, simultaneous, and concurrent tasks. *Parallel tasks* can informally execute in parallel. *Simultaneous tasks* execute at the same time. And *concurrent tasks* are independent of each other (namely, none of these tasks consumes any data element that is produced by another concurrent task); hence, concurrent tasks can execute in any order, simultaneously or sequentially.

Task concurrency is defined by the task graph. If there exists a transitive path from task $t_i$ to task $t_j$, implying that task $t_i$ must complete before task $t_j$ is ready and can start execution, tasks $t_i$ and $t_j$ are not concurrent. Otherwise, these tasks are concurrent. These definitions suggest that the task graph may be partitioned into concurrent sets, where each set contains tasks that are mutually concurrent. Note that this suggestion is not always true, but a more complete treatment of this topic lies beyond the scope of this paper. Further, cycles in the graph usually imply repetitions or pipelined data generations and may need to be broken during the analysis process.

Given the concurrent tasks analysis, we may be able to formally verify that the code adheres to PRAM CREW parallel processing using shared memory. First, we identify the memory addresses of all read and write operations. To enable this identification, all shared memory addresses should be derivable at compile time, no data-dependent shared memory addresses (such as data-dependent pointers) are allowed, and all dynamic memory allocations (by *malloc*) should employ only predictable addresses.

To verify Exclusive Write (EW), all write operations should be checked. If task $t_i$ writes into variable in address $A$, then no concurrent task $t_k$ is allowed to access $A$ (neither read nor write).

To verify Concurrent Read (CR), all read operations should be checked. If task $t_i$ reads a variable in address $A$, then no concurrent task $t_k$ is allowed to write into $A$.

A RC64 program that is thus verified does not need any cache coherency. Its execution is provably correct. Note that formal verification is required because nothing else prevents programming errors that violate the rules, and these errors may evade any simulation and functional verification efforts.

# 10. RC64 Memory Model

The standard memory organization employed by languages such as C in a single core computer running a variety of operating systems is shown in Figure 15. Coding for RC64 shared memory manycore should be perceived as similar as possible to serial programming, and the shared memory organization should resemble a single core computer. Figure 15, on the right, illustrates that the only unshared memory part is the stack. Each core maintains its own stack. Although the stack address sub-space is the same for all cores, the contents are different. This memory model is exemplified in the next section.

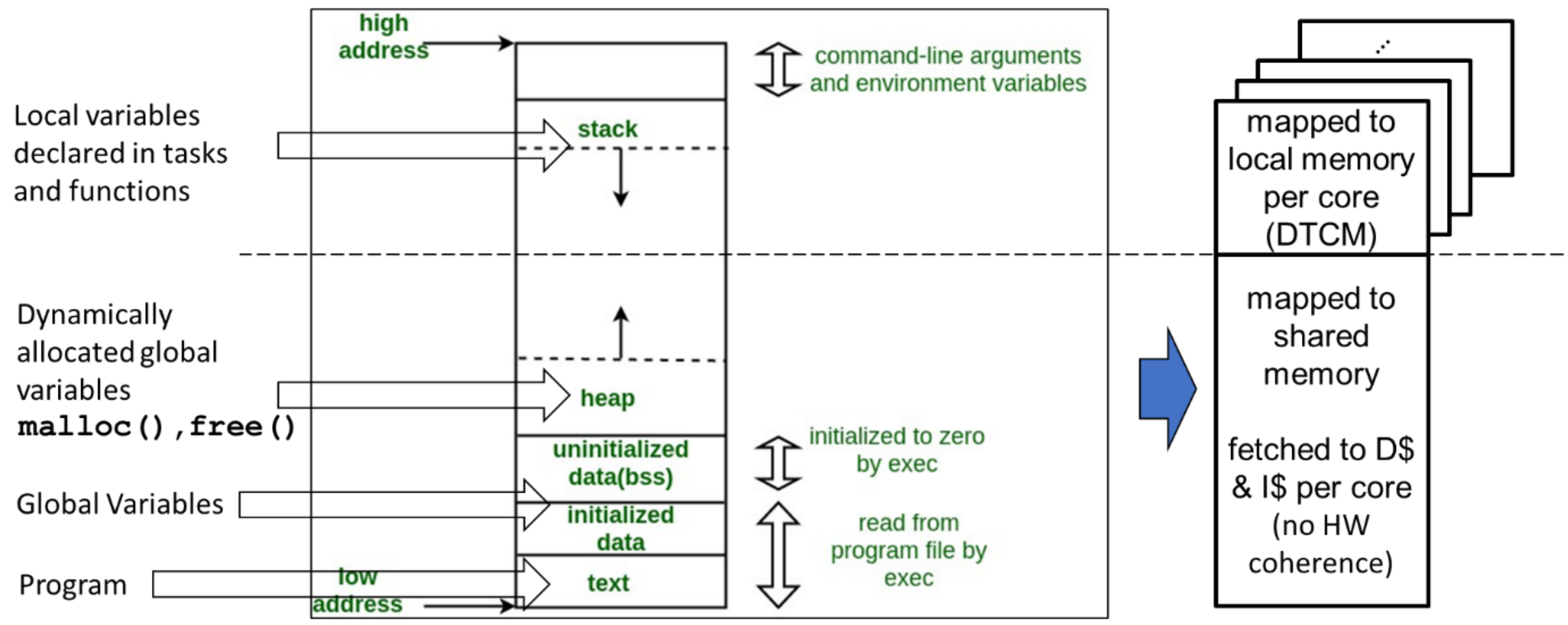

***Figure 15. Left: standard memory model for single core programs. Right: RC64 memory model. The stack is mapped to the local private memory of each core ("data tightly coupled memory", DTCM), and is not cached. Other parts of the address space are mapped to shared memory and can be cached. No cache coherency mechanism is needed.***

# 11. RC64 Matrix Multiplication Algorithm

A simple linear algebra program demonstrates the programming model, the memory model, and the RC64 design. These concepts are applicable to other algorithms, such as deep learning.

In general, $C = A \times B$ where $A, B$ and $C$ are $N \times N$ matrices. In more detail, $C_{i,j} = \sum_m A_{i,m} \times B_{m,j}$. Each result element $C_{i,j}$ is independent of all other result elements and can be computed by a separate instance of a duplicable task. The different instances follow CREW sharing. Sample code is shown on the left side in Figure 16. The scheduler assigns a unique ID to each instance, and the instance computes its row and column indices from that ID. The instance performs dot product and writes the result to shared memory.

The shared matrices are declared globally outside the scope of the tasks. Tasks are coded as C functions. Task codes are never invoked by other functions, and are initiated only by the hardware scheduler. Regular tasks do not take any parameters, but may produce a return code (enabling choice in the task graph). Duplicable tasks take a single argument, the instance ID, and must not generate any return code. Recall that no affinity is implied between an instance ID and a core number—any core can execute any instance.

```
#define M 100
float  A[M][M], B[M][M], C[M][M];

int  mm_start()                              REGULAR
{   int i,j;
    for (i=0; i< M; i++)
        for (j=0; j< M; j++)
            { A[i][j] = 13;  B[i][j] = 9;  }
}
void  mm (unsigned int id)                   DUPLICABLE
{   int  i,j,m;  float sum = 0;
    i = id % M;      j = id / M;
    for (m=0; m < M; m++)
        sum += A[i][m]*B[m][j];
    C[i][j]=sum;
}
int  mm_end ()                               REGULAR
{  printf("finished mm\n");  }
```

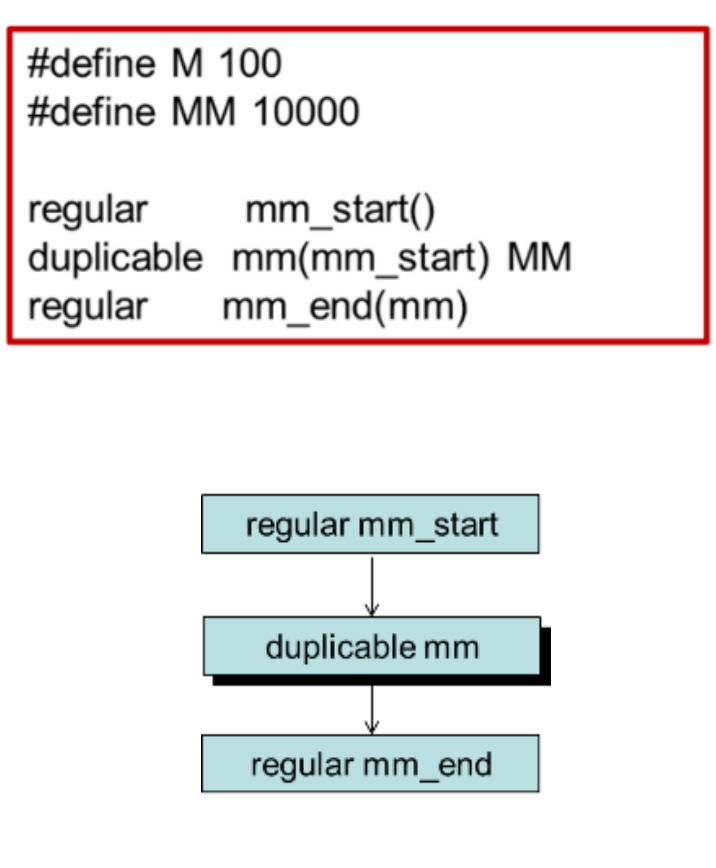

***Figure 16. Left: matrix multiplication code, consisting of two regular and one duplicable tasks. Right: Task graph for matrix multiplication.***

The corresponding task graph is shown on the right in Figure 16, in both text and graph form. The quota for the duplicable task is specified in the task graph. Alternatively, it can be set by the code, in any task that is precedent to it.

The memory model is shown in Figure 17. Shared memory holds the three matrices and the program code. Private variables, automatically generated in the stack of each executing task, end up in DTCM private memory of each core, and are valid only during the execution of an instance. The DTCM is invalidated between successive executions of instances on the same core. I$ fetches the code, and may stay 'warm' between successive instances. D$ fetches the relevant row and column per each instance. D$ may also stay warm between instances, but this rule is opportunistic and is neither managed nor guaranteed to be useful.

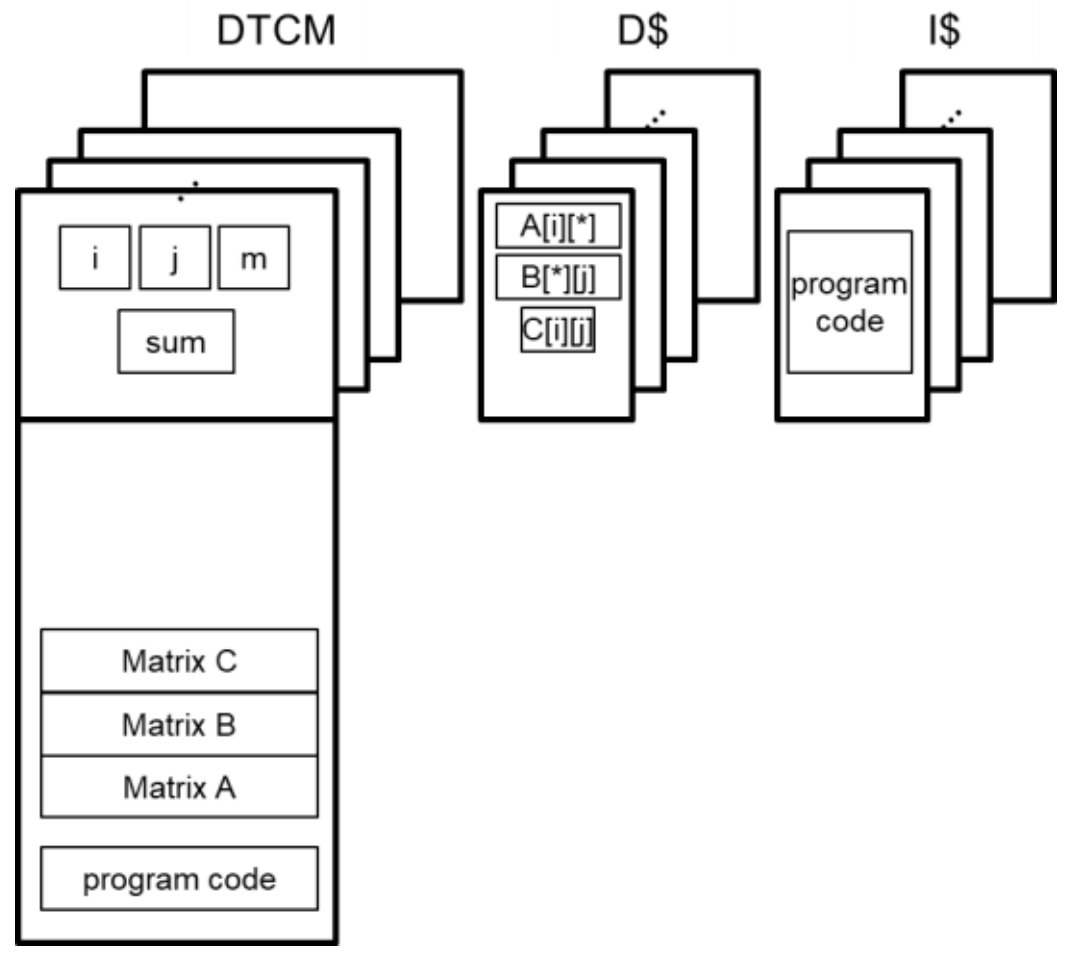

***Figure 17. RC64 memory model for matrix multiplication***

Pipelined matrix multiplication may be programmed as three stages: read matrices *A* and *B*, compute *C*, and write *C*. A 'double buffering' concept would compute generation $C_i$ while outputting $C_{i-1}$ and reading the next set of matrices $A_{i+1}$, $B_{i+1}$. A generation index is toggled 0,1 by task *L* in Figure 18. The shared memory needs to accommodate six matrices (in contrast to the three matrices mentioned above). The code, task graph, and memory model are shown in Figure 19.

Figure 18. Task graphs for pipelined matrix multiplication (double buffers), showing explicit generation indices (left) and a more succinct style (right). Task L toggles the generation indices. (e) indicates a hardware event (completion of input or output operation).

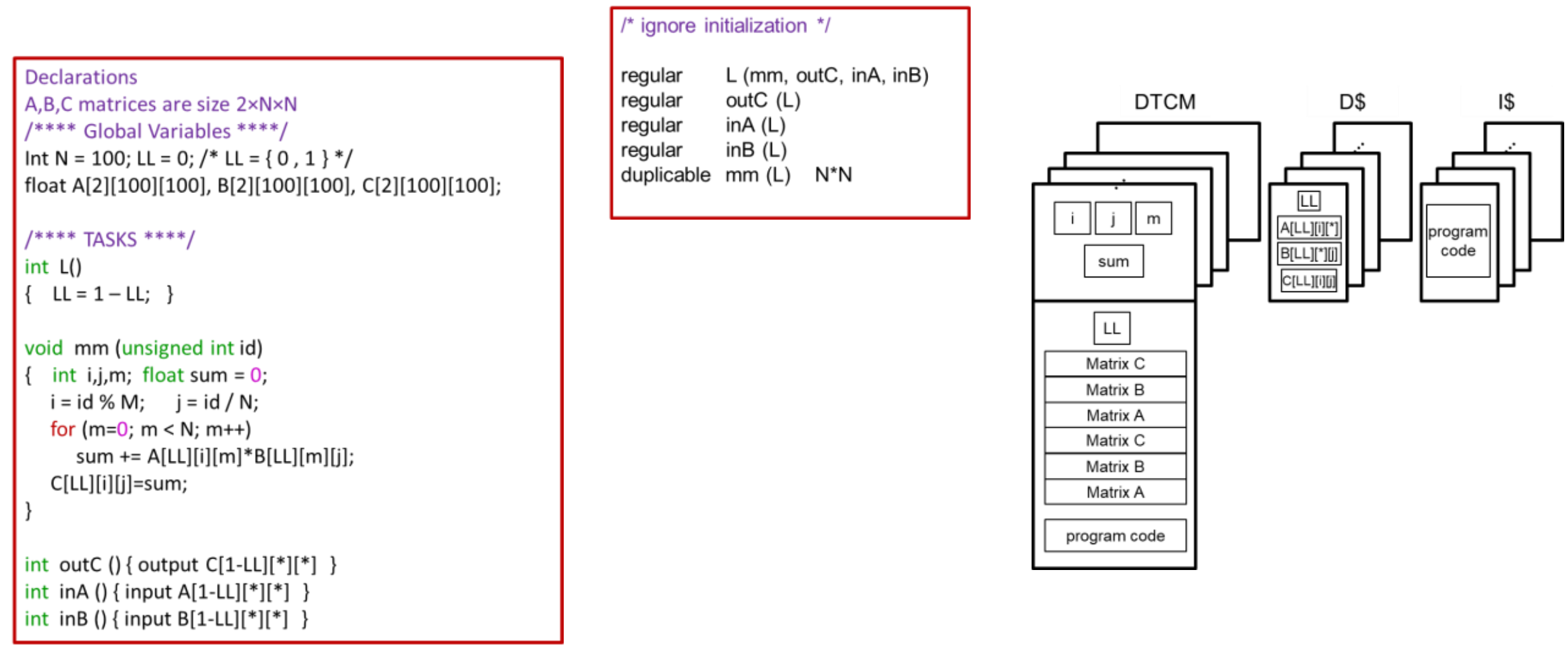

Figure 19. Code, task graph and memory model for pipelined matrix multiplication. Row order in the task graph ensures that three cores briefly execute outC(), inA(), inB() to issue DMA activities. These cores 'quickly' join other cores in executing instances of mm().

## 12. Ramon.Space Future Manycore

The follow-up to RC64 is only partly designed. As shown in Figure 20, we consider breaking the manycore down into multiple GALS domains [52]. GALS system design implies variable latency of accessing shared memory, depending on distance (counted by the number of traversed GALS modules). With RC64, it became evident that variable delays on accessing shared memory have almost no effect on performance, especially when most accesses fetch cache lines and pre-fetch methods are employed. Hence, GALS does not really degrade performance. GALS architecture also does not lower power consumption. So why bother? The main arguments are the reduction and modularization in the physical design effort and the alleviation of the global timing closure challenge. In fact, global timing closure caused increased expenses and delays in taping out RC64.

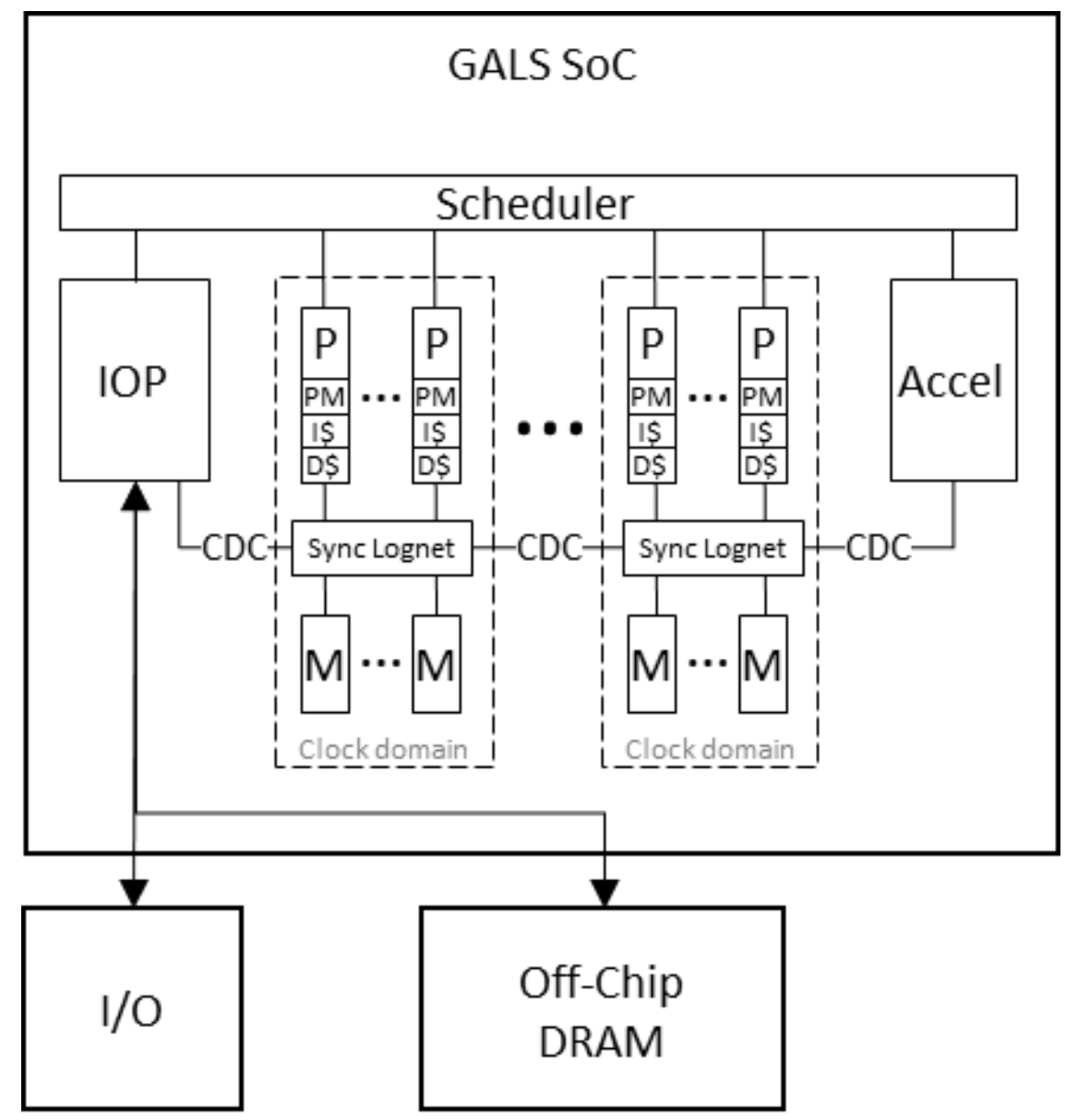

*Figure 20. Conceptual architecture for the evolution of RC64.*

# 13. Gratitude[2]

I feel privileged and happy to have spent a long time contributing to parallel processing. I also feel grateful for the opportunity to learn asynchronous design and apply my deeper understanding to improve parallel computers.

# References

[1] Bayer, Nimrod. "A hardware-synchronized/scheduled multiprocessor model." *Technion–Israel Institute of Technology, Thesis, English abstract online, http://webee. technion. ac. il/~ ran/papers/NimrodBayerMScThesisAbstract1989. pdf January* (1989).

[2] Plurality Ltd. HAL architecture is described in https://cordis.europa.eu/docs/projects/cnect/6/248776/080/deliverables/001-PRO3DD143dmemoryhierarchiesv32.pdf (accessed July 2026).

[3] Ginosar, Ran, Peleg Aviely, Tsvika Israeli, and Henri Meirov. "RC64: High performance rad-hard manycore." In 2016 IEEE Aerospace Conference, pp. 1-9. IEEE, 2016.

[4] Aviely, Peleg, Olga Radovsky, and Ran Ginosar. "DVB-S2 software defined radio modem on the RC64 manycore DSP." 2016 IEEE Aerospace Conference. IEEE, 2016.

[5] Stockmeyer, Larry, and Uzi Vishkin. "Simulation of parallel random access machines by circuits." *SIAM Journal on Computing* 13.2 (1984): 409-422.

[6] Spars, Jens, and Steve Furber. *Principles asynchronous circuit design*. Kluwer Academic Publishers, 2002.

[7] Ullman, Jeffrey D. *Principles of database systems*. Galgotia publications, 1983.

[8] Kleinrock, L. (1974). *Queueing systems (two volumes)*. Wiley.

[9] Russell, Richard M. "The CRAY-1 computer system." *Communications of the ACM* 21.1 (1978): 63-72.

[10] Ibbett, R. N., and N. P. Topham. "The CDC Series." *Architecture of High Performance Computers: Volume I Uniprocessors and vector processors*. New York, NY: Springer New York, 1989. 156-179.

[11] Owens, John D., David Luebke, Naga Govindaraju, Mark Harris, Jens Krüger, Aaron E. Lefohn, and Timothy J. Purcell. "A survey of general-purpose computation on graphics hardware." In *Computer graphics forum*, vol. 26, no. 1, pp. 80-113. Oxford, UK: Blackwell Publishing Ltd, 2007.

[12] Patterson, David, and Andrew Waterman. "SIMD instructions considered harmful." *ACM SIGARCH.[Online]. Available: https://www. sigarch. org/simd-instructions-considered-harmful* (2017).

[13] Tang, Ju-ho, Edward S. Davidson, and Johau Tong. "Polycyclic vector scheduling vs. chaining on 1-port vector supercomputers." *Supercomputing'88: Proceedings of the 1988 ACM/IEEE Conference on Supercomputing, Vol. I*. IEEE, 1988.

[14] Manolis Katevenis Reduced Instruction Set: Computer Architectures for VLSI. ACM doctoral dissertation awards, MIT Press, 1985

[15] Horowitz, and Zorat. "Divide-and-conquer for parallel processing." *IEEE Transactions on Computers* 100.6 (1983): 582-585.

[16] Boberg. "Proposed microcomputer system IEEE-796 bus standard." *Computer* 13.10 (1980): 89-105.

[17] Arden, and Ginosar. "MP/C: A multiprocessor/computer architecture." *IEEE Transactions on Computers* 100.5 (1982): 455-473.

[18] Bucher, Ingrid Y., and James W. Moore. *Comparative performance evaluation of two supercomputers: CDC Cyber-205 and CRI Cray-1*. No. LA-UR-81-1977; CONF-811202-2. Los Alamos National Lab., NM (USA), 1980.

[19] Dennis, Jack B., and David P. Misunas. "A preliminary architecture for a basic data-flow processor." *Proceedings of the 2nd annual symposium on Computer architecture*. 1974.

[20] Kung, Hsiang Tsung, and Charles E. Leiserson. "Systolic arrays (for VLSI)." *Sparse Matrix Proceedings 1978*. Vol. 1. Philadelphia, PA, USA: SIAM, 1979.

[21] Bechtolsheim, Andrew. *The SUN workstation architecture*. Stanford University, 1982.

[22] https://www.nobelprize.org/prizes/physics/1978/ceremony-speech/

[23] Cook, Stephen A., and Robert A. Reckhow. "Time-bounded random access machines." *Proceedings of the fourth annual ACM symposium on Theory of computing*. 1972.

[24] Vishkin, Uzi. *Synchronized parallel computation*. Technion-Israel Institute of Technology, Faculty of Computer Science PhD Thesis, 1981.

[25] Schwartz, Jacob T. "Ultracomputers." *ACM Transactions on Programming Languages and Systems (TOPLAS)* 2, no. 4 (1980): 484-521.

[26] Gottlieb, Grishman, Kruskal, McAuliffe, Rudolph, and Snir. "The NYU ultracomputer—Designing an MIMD shared memory parallel computer." *IEEE Transactions on computers* 100, no. 2 (1983): 175-189.

[27] Pfister, G. F., W. C. Brantley, D. A. George, S. L. Harvey, W. J. Kleinfelder, K. P. McAuliffe, E. A. Melton, V. A. Norton, and J. Weiss. "The IBM research parallel processor prototype (RP3): Introduction and architecture." In *1985 Int'l. Conf. Parallel Processing*, pp. 764-771. 1985.

[28] Lipovski, G. J., & Vaughan, P. (1988). A fetch-and-op implementation for parallel computers. *ACM SIGARCH Computer Architecture News*, *16*(2), 384-392.

[29] Snir, Marc. *MPI--the Complete Reference: the MPI core*. Vol. 1. MIT press, 1998.

[30] W. C. Athas and C. L. Seitz, "Multicomputers: message-passing concurrent computers," in *Computer*, vol. 21, no. 8, pp. 9-24, Aug. 1988.

[31] https://axautikgroupllc.substack.com/p/the-nivida-bluefield-dpu-from-acquisition (online, retrieved July 2026).

[32] Mittal S. A survey on evaluating and optimizing performance of Intel Xeon Phi. *Concurrency Computat Pract Exper*. Wiley, 2020.

[33] Amdahl, Gene M. "Computer architecture and Amdahl's law." *Computer* 46, no. 12 (2013): 38-46.

[34] Gustafson, John L. "Reevaluating Amdahl's law." *Communications of the ACM* 31.5 (1988): 532-533.

[35] Nickolls, John, Ian Buck, Michael Garland, and Kevin Skadron. "Scalable parallel programming with CUDA: Is CUDA the parallel programming model that application developers have been waiting for?" *Queue* 6, no. 2 (2008): 40-53.

[36] Leiserson, Charles, and Aske Plaat. "Programming parallel applications in Cilk." *SINEWS: SIAM News* 31, no. 4 (1998): 6-7.

[37] Naishlos, Dorit, Joseph Nuzman, Chau-Wen Tseng, and Uzi Vishkin. "Evaluating the XMT parallel programming model." In *International Workshop on High-Level Parallel Programming Models and Supportive Environments*, pp. 95-108. Berlin, Heidelberg: Springer Berlin Heidelberg, 2001.

[38] Stockmeyer, Larry, and Uzi Vishkin. "Simulation of parallel random access machines by circuits." *SIAM Journal on Computing* 13, no. 2 (1984): 409-422.

[39] O. Aumage et al., Task-Based Performance Portability in HPC, ETP4HPC White Paper, 2021, doi 10.5281/zenodo.5549731

[40] Mead, Carver, and Lynn Conway. *Introduction to VLSI systems*. 1980.

[41] David, Ilana, Ran Ginosar, and Michael Yoeli. "Implementing sequential machines as self-timed circuits." *IEEE Transactions on Computers* 41.1 (1992): 12-17.

[42] David, Ilana, Ran Ginosar, and Michael Yoeli. "An efficient implementation of Boolean functions as self-timed circuits." *IEEE transactions on computers* 41.1 (1992): 2-11.

[43] Martin, A.J. (1990). The Limitations to Delay-Insensitivity in Asynchronous Circuits. In: Feijen, W.H.J., van Gasteren, A.J.M., Gries, D., Misra, J. (eds) Beauty Is Our Business. Texts and Monographs in Computer Science. Springer, New York, NY.  Seminar paper on async logic by alain martin about delay insensitive logic

[44] Jo C. Ebergen: A Formal Approach to Designing Delay-Insensitive Circuits. Distributed Comput. 5: 107-119 (1991)

[45] Beerel, Peter A., Recep O. Ozdag, and Marcos Ferretti. *A designer's guide to asynchronous VLSI*. Cambridge University Press, 2010.

[46] J. Cortadella, M. Kishinevsky, A. Kondratyev, L. Lavagno, A. Yakovlev, “Logic Synthesis for Asynchronous Controllers and Interfaces,” 2002 Springer

[47] Stevens, S. Rotem, R. Ginosar, P. Beerel, C.J. Myers, K.Y. Yun, R. Kol, C. Dike and M. Roncken, “An Asynchronous Instruction Length Decoder,” IEEE Journal of Solid State Circuits, 36(2), pp. 217-228, Feb. 2001

[48] K. S. Stevens, R. Ginosar and S. Rotem, "Relative timing [asynchronous design]," in *IEEE Transactions on Very Large Scale Integration (VLSI) Systems*, vol. 11, no. 1, pp. 129-140, Feb. 2003

[49] Furber, S. B., Garside, J. D., Riocreux, P., Temple, S., Day, P., Liu, J., & Paver, N. C. (1999). AMULET2e: An asynchronous embedded controller. *Proceedings of the IEEE*, *87*(2), 243-256.

[50] R. Kol and R. Ginosar, “KIN--An Asynchronous Processor.” 12th ACM International Conference on Supercomputing (ICS’98), Jul. 1998.

[51] R. Dobkin, R. Ginosar, A. Kolodny, QNoC Asynchronous Router, Integration, the VLSI Journal, 42(2):103-115, 2009.

[52] R. Dobkin, R. Ginosar and C. Sotiriou, High Rate Data Synchronization in GALS SoCs, IEEE Trans. on VLSI, 14(10):1063-1074, Oct. 2006.

[53] Ginosar, Ran, et al. “Ramon space RC64-based AI/ML inference engine.” ESA European Workshop on On-Board Data Processing. 2021

[54] Ginosar, Ran, et al. "Machine learning space applications using RC64 rad hard manycore processor." 2023 European Data Handling & Data Processing Conference (EDHPC). IEEE, 2023.

[55] Ghiglino, Pablo, et al. "Ramon. Space RC64 and NuStream with Klepsydra AI: Rad-hard High Performance On-Board AI." 2025 European Data Handling & Data Processing Conference (EDHPC). IEEE, 2025.

[56] Ghiglino, Pablo, et al. "Performance Benchmark of Ramon Space RC64 and NuStream with Klepsydra AI: Towards a Full Rad-Hard High Performance SW/HW Solution." 2025 IEEE Space Computing Conference (SCC). IEEE, 2025.

[57] Ginosar, Ran, et al. "Vision based lunar landing using RC64 rad hard DSP/ML manycore processor." 2023 European Data Handling & Data Processing Conference (EDHPC). IEEE, 2023.